\documentclass[fleqn,twoside]{article}
\usepackage{espcrc2}
\usepackage{epsfig}
\newcommand{\lsim}{\stackrel{<}{_\sim}}

\usepackage{graphicx}
\usepackage[figuresright]{rotating}

\newcommand{\AmS}{{\protect\the\textfont2
  A\kern-.1667em\lower.5ex\hbox{M}\kern-.125emS}}

\title{physics reach of rare $b$-decays 
\thanks{Invited talk at the 5th International
Conference on Hyperons, Charm and Beauty Hadrons (BEACH 2002),
University of British Columbia, Vancouver, June 25-29, 2002.}
\hspace{5.5cm} 
{\large{SLAC-PUB-9277\\
 \hspace{13cm} July 2002}}
}

\author{Gudrun Hiller \address{
Stanford Linear Accelerator Center, Stanford University, Stanford, 
CA 94309, USA }
        \thanks{Work
supported by the Department of Energy, Contract
DE-AC03-76SF00515}}
       
\begin{document}

\begin{abstract}
I discuss the theoretical and empirical status of 
$b \to s \gamma$,
$b \to s \ell^+ \ell^-$ decays, as well as their future prospects.
I emphasize those observables in rare $b$-decays which can potentially
establish new physics and distinguish between extensions of the
Standard Model.
I briefly review current models of electroweak symmetry breaking, 
all of which can carry interesting flavor characteristics 
accessible with $b$-physics experiments.

\vspace{1pc}
\end{abstract}

\maketitle

\section{$b \to s \gamma$ STATUS}

Rare radiative $b \to s \gamma$ decays are both theoretically and
experimentally well studied and reached attention as an important
constraint on extensions of the Standard Model (SM). 
The branching ratio is known to
NLO (see e.g.~\cite{misiak}) and depends only at 2-loop (the 1-loop
matrix element vanishes for an on-shell photon) on the charm mass.
But even it appears only at 2-loop, different choices of $m_c$ do
numerically matter \cite{misiak}
\begin{equation}
\label{eq:pole}
{\cal{B}}(B \to X_s \gamma)_{SM}=(3.35 \pm 0.30) \cdot 10^{-4}
\end{equation}
if the pole mass $m_c^{pole}/m_b =0.29 \pm 0.02$ is used and
\begin{equation}
\label{eq:msbar}
{\cal{B}}(B \to X_s \gamma)_{SM}=(3.73 \pm 0.30) \cdot 10^{-4}
\end{equation}
for the Ms-bar mass renormalized at a scale $\sim m_b$
$\bar m_c(\mu)/m_b =0.22 \pm 0.04$.
The difference between
$m_c^{pole}$ and  $\bar m_c(\mu)$ is a higher order in $\alpha_s$ issue. 
In the absence of a 3-loop NNLO calculation to identify the correct 
$m_c$ prescription 
we combine the above branching ratios, inflate errors and obtain
\begin{equation}
{\cal{B}}(B \to X_s \gamma)_{SM}=(3.54 \pm 0.49) \cdot 10^{-4}
\end{equation}
Comparison with the data by Cleo, Aleph and Belle (Ref.~[3-5] in \cite{aghl})
\begin{equation}
{\cal{B}}(B \to X_s \gamma)_{world ave}=(3.22 \pm 0.40) \cdot
10^{-4}
\end{equation} 
shows that the theory error exceeds the experimental one and
unless there is progress in theory 
(or the experimental central value moves a lot), we cannot
establish new physics (NP) with the $B \to X_s \gamma$ branching ratio alone.

\begin{figure}[htb]
\vskip -0.2truein
\centerline{\epsfysize=2.0in
{\epsffile{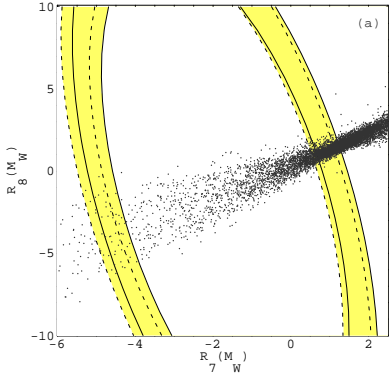}}}
\vskip -0.4truein
\caption{Constraints from ${\cal{B}}(b \to s \gamma)$ @ $90\%$C.L. on
$R_{7,8}(m_W)$. Figure taken from \cite{aghl}.}
\label{fig:r7r8}
\end{figure}

Model independent constraints from ${\cal{B}}(B \to X_s \gamma)$
have been obtained in the effective Hamiltonian theory
${\cal{H}}_{eff}= -4 G_F/ \sqrt{2} V_{tb} V_{ts}^* \sum C_i O_i$ with
effective vertices $O_i$
and Wilson coefficients $C_i$ \cite{aghl}.
Important here are the operators
$O_7\propto \bar{s}_L \sigma_{\mu \nu} b_R F^{\mu \nu}$ and
$O_8\propto \bar{s}_L \sigma_{\mu \nu} b_R G^{\mu \nu}$.
The LO branching ratio
${\cal{B}}(B \to X_s \gamma)_{LO} \propto |C_7(m_b)|^2$ fixes the
modulo of $C_7$, which illustrates that one can measure the $C_i$.
Constraints on $C_{7,8}$ have been worked out at NLO in terms of the ratios
$R_i(\mu) \equiv (C_i^{SM}(\mu)+C_i^{NP}(\mu))/C_i^{SM}(\mu)$ \cite{aghl}.
The result is shown in Fig.~\ref{fig:r7r8} at $\mu=m_W$. The bands are 
the allowed regions,
the SM is $R_{7,8}=1$. The solid (dashed) lines denote the bound using
a pole (Ms-bar) mass prescription for the charm quark.
Future expectations are such that
by 2005 the $B$-factories have collected $500 fb^{-1}$ and
measured the ${b \to s \gamma}$ branching ratio precisely
$\sigma(stat,sys)= 1.8 \%,3 \%$ \cite{Eigen:2001tb}.
Hence, the 2 bands would be very
narrow, approximately of the size of the difference between the
solid and dashed lines given by todays dominant theory error, $m_c$.

The scatter plot results from a scan over the minimal supersymmetric
model (MSSM) parameter space with
minimal flavor violation (MFV). This is defined as no more flavor
violation than in the SM, i.e.~in the yukawas $Y$. 
In supersymmetry (SUSY) this is a condition on the SUSY breaking and
enforces proportionality of A-terms $\sim Y$ and degeneracy of squark
masses.
MFV examples are gauge and anomaly mediation. 
The parameters of the MSSM-MFV scan obey
$m_{\tilde t},m_{\chi} > 90$ GeV, $m_{\tilde \nu} > 50$ GeV,
$m_{H^\pm} > 78.6$ GeV,   
$|\mu|,M_2 < 1$ TeV,
$2.3 < \tan \beta < 50$ and stop mixing angle
$|\Theta_{\tilde t}|< \pi/2$.
The solution with the sign of $C_7$ flipped w.r.t.~the SM needs a
large chargino-stop contribution, since both SM and charged Higgs ones
interfere constructively, thus large $\tan \beta$ and/or a light stop is 
required.
Fig.~\ref{fig:r7r8} demonstrates that ${\cal{B}}(b \to s \gamma)$ 
data cut out many points and both
branches are allowed and can be reached by NP.
To distinguish them we need additional constraints 
such as from $b \to s \ell^+ \ell^-$ decays.

\section{$b \to s \ell^+ \ell^-$ OPPORTUNITIES}
Rare $b \to s \ell^+ \ell^-$ decays have besides the $\gamma$ penguin
known to $b \to s \gamma$ decays additional structures: the $Z$ penguin and
the box diagram, which are encoded in the operators
$O_9 \propto (\bar{s}_L \gamma_\mu b_L )
(\bar{\ell}  \gamma^\mu \ell)$ and 
$O_{10} \propto (\bar{s}_L  \gamma_\mu b_L) 
( \bar{\ell} \gamma^\mu \gamma_5 \ell)$.
The first mode mediated by $b \to s \ell^+ \ell^-$, exclusive
$B \to K \ell^+ \ell^-$ decays has recently
been observed by Belle \cite{Abe:2001dh}, and also by Babar \cite{babarKll}
\begin{eqnarray}
{\cal{B}}(B \to K \ell^+ \ell^-)_{\cite{Abe:2001dh}}
\!\!&\!=\!& \!\!0.75^{+0.25}_{-0.21} \pm 0.09  \cdot 
10^{-6} \\
{\cal{B}}(B \to K \ell^+ \ell^-)_{\cite{babarKll}}
\!\!&\!=\!&\!\!0.84^{+0.30+0.10}_{-0.24-0.18} \cdot 10^{-6}
\end{eqnarray}
with rates comparable to the
SM \cite{aghl}
\begin{equation}
{\cal{B}}(B \to K \ell^+ \ell^-)_{SM}=0.35 \pm 0.12  \cdot 10^{-6}
\label{eq:KllSM}
\end{equation}
The SM calculation \cite{aghl} 
is performed at NNLO (see below the discussion of
inclusive decays) assuming factorization. 
Corrections from
spectator interactions have been ignored, because this is a sub leading 
effect compared to the dominant source of theoretical uncertainty, 
i.e.~the form factors.
The reduction of ${\cal{B}}(B \to K \ell^+ \ell^-)$ 
with respect to earlier calculations obtained
at NLO by $39 \%$ (central values) \cite{ABHH} has 2 sources, first
the effects of the NNLO calculation 
which are also active in the inclusive decays and
secondly the lower central value of form factors, as suggested by related
analyses in $B \to K^* \gamma$ decays \cite{aghl}.
To be specific, the minimum set of form factors from light cone QCD sum
rules from Ref.~\cite{ABHH} plus a $\pm 15 \%$ error has been
used to obtain Eq.~(\ref{eq:KllSM}).

\begin{table*}[htb]
\caption{Experimental status of inclusive $B \to X_s \ell^+ \ell^-$
decays.}
\label{tab:Xsll}
\renewcommand{\arraystretch}{1.11}
         \begin{center}
         \begin{tabular}{c|cc|c}
\hline
mode & branch fraction Belle'02 \cite{bellesenyo} & 
signif. & upper bound Belle'01 \cite{Abe:2001qh}\\
\hline
$B \to X_s \mu^+ \mu^-$ &$8.9^{+2.3+1.6}_{-2.1-1.7} \cdot
         10^{-6}$
& $4.4 \sigma$ & $ < 19.1  \cdot 10^{-6}$ 
@ $90 \%$ C.L. \\
$B \to X_s e^+ e^- $& $5.1^{+2.6+1.3}_{-2.4-1.2} \cdot
         10^{-6}$  & $2.1 \sigma$ & $ < 10.1  \cdot 10^{-6}$  @ $90 \%$
C.L.\\
\hline
         \end{tabular}
         \end{center}
\end{table*}

Inclusive $b \to s \ell^+ \ell^-$ decays are known to NNLO accuracy for
low dilepton mass 
\cite{BobethNNLO}-\cite{Asatryan:2002iy}. 
The effective coefficients
$C_i^{eff}=\left[1+\frac{\alpha_s}{\pi} \omega_i(\hat s)\right] A_i + 
\frac{\alpha_s}{4 \pi}  C_j F_{ij}(\hat s)$ include
virtual $\alpha_s$-corrections in the functions 
$F_{ij}$ and bremsstrahlung and $\alpha_s$-corrections to
the matrix elements $<O_i>$ in the $\omega_{i}$.
The $b\to s \ell^+\ell^-$ decay rate can then be written as a function
of the normalized dilepton mass $\hat s =q^2/m_b^2$ as
\cite{aghl}
\begin{eqnarray}
    \frac{d\Gamma}{d\hat s} \!\!\! &\!\!\!  \sim \!\!\! & \!\!\!
    (1-\hat s)^2 \left[ \left (1+2\hat s\right)
    ( |C_9^{eff} |^2+
     |C_{10}^{eff} |^2  ) f_1( \hat s) 
\right. \nonumber \\
    &+& \left. 4\left(1+2/\hat s\right)
    | C_7^{eff} |^2  f_2( \hat s) \right. \nonumber \\
    &+& \left.
    12 \mbox{Re} ( C_7^{eff}
    C_9^{eff*} ) f_3( \hat s) + f_c(\hat s) \right]  
\end{eqnarray}
including $1/m_c$ \cite{Buchalla:1997ky} and
$1/m_b$ \cite{Ali:1996bm} corrections in the functions $f_c$ and $f_{1,2,3}$.
Experimental information on inclusive $B \to X_s \ell^+ \ell^-$ decays
from Belle
\cite{Abe:2001qh,bellesenyo} is compiled in Tab.~\ref{tab:Xsll} and
is consistent with the SM branching ratios \cite{aghl}
\begin{eqnarray}
{\cal{B}}(B \to X_s e^+ e^-)\!\!\!\!&\!\!=\!\!&\!\!\!\!6.89 \pm  0.37 \pm
0.25 \pm 0.91 
\! \cdot \!10^{-6} \nonumber\\
{\cal{B}}(B \to X_s \mu^+ \! \mu^-)\!\!\!\!&\!\!=\!\!&\!\!\!\!
4.15 \pm  0.27 \pm 0.21  \pm 
0.62  \! \cdot \! 10^{-6} \nonumber
\end{eqnarray}
where the errors correspond to 
varying $m_b/2 <\mu < 2 m_b$, $m_t^{pole}=(173.8 \pm 5)$ GeV and
$m_c/m_b =0.29 \pm 0.04$, respectively.
Adding them in quadrature the total errors are estimated as
$\delta {\cal{B}}_{X_see} = \pm 15 \%$ and
$\delta {\cal{B}}_{X_s \mu \mu } = \pm 17 \%$.
Going from NLO to NNLO decreases 
${\cal{B}}(B \to X_s \ell^+ \ell^-)$ 
by $12 \%$ for dielectrons and $20 \%$ for dimuons.
The full NNLO calculation, i.e.~the functions 
$F_{ij}$ are only available for $\hat s < 0.25$ below the $c \bar c$ 
threshold.
The above branching ratios are 
obtained from naive extrapolation of the $F_{ij}$, which gives a
spectrum that is 
well approximated for all $\hat s$ by the partially NNLO one with
$F_{ij} \equiv 0$ for $\mu \simeq m_b/2$ \cite{aghl}.
Contributions from charmonium vector mesons via
$b \to s V \to s \ell^+ \ell^-$ should be removed from the data by
cuts in the dilepton mass around $q^2=m_{J/\Psi}^2,m_{\Psi^\prime}^2$ 
\cite{ABHH,Ali:1996bm}.

The biggest source of theory uncertainty in the 
$B \to X_s \ell^+ \ell^-$ branching ratios is $m_c$ \cite{aghl}. 
It appears already at 1-loop and is conservatively varied as
$m_c/m_b=0.29 \pm 0.04$ \cite{Asatryan:2002iy}.
Study of the
parametric dependence of the decay rates on $z=m_c/m_b$ 
such that $\epsilon$ denotes the percental correction to $\Gamma$ if
$z$ changes from $0.29$ by $-0.04$ as
\begin{eqnarray}
\nonumber
\frac{\Gamma(z)-\Gamma(0.29)}{\Gamma(0.29)} 
\approx \frac{0.29-z}{0.04}  \epsilon +{\cal{O}}((0.29-z)^2)
\end{eqnarray}
yields
$\epsilon=2 \%,6 \%,16 \%$ for $b \to s \ell^+ \ell^-$, $b \to s \gamma$,
$b \to c \ell^- \bar{\nu}_\ell$ decays, respectively.
This explains the shift of $11 \%$ in ${\cal{B}}(b \to s \gamma)$ 
see Eqs.~(\ref{eq:pole}),(\ref{eq:msbar}) when
going from the pole mass to the Ms-bar charm mass.
Further, 
the bulk of the $m_c$ dependence in ${\cal{B}}(b \to s \ell^+ \ell^-)$ 
does not result from the 
$b \to s \ell^+ \ell^-$ decay rate, but from the 
normalization to $\Gamma(b \to c \ell^- \bar{\nu}_\ell)$, which
is employed to remove the $m_b^5$ dependence. 
The decay rates of exclusive $B \to (K,K^*) \ell^+ \ell^-$ 
decays are normalized to the $B$-lifetime, hence there the
theoretical error due to $m_c$ is small.
By 2005 $B$-factories are  expected to have collected a 
few hundred events of both $b \to s e^+ e^-,s \mu^+ \mu^-$ decays
and measured the branching ratios with 
$\sigma(stat,sys)_{e^+ e^-} \simeq 7 \%,7 \%$ 
and $\sigma(stat,sys)_{\mu^+ \mu^-} \simeq 9 \%,12 \%$
\cite{Eigen:2001tb}.

\begin{figure}[htb]
\epsfig{file=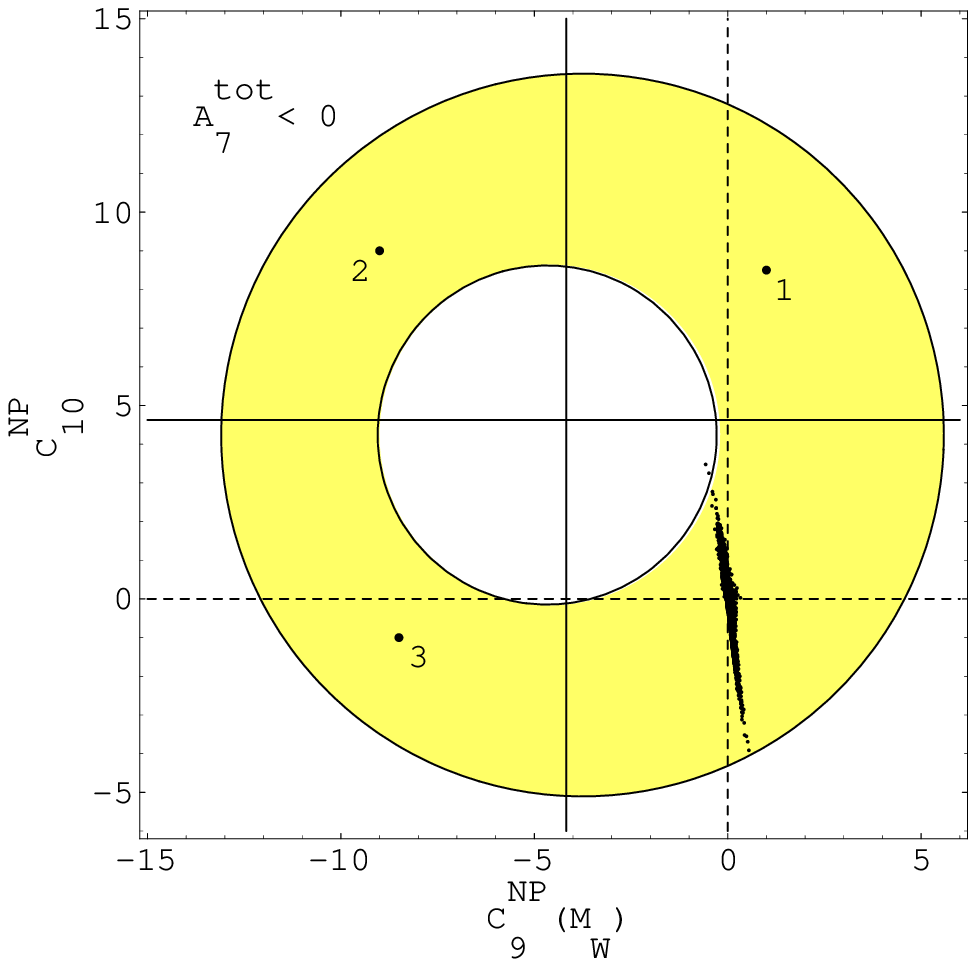,width=0.51\linewidth}
\hspace*{-0.3cm}
\epsfig{file=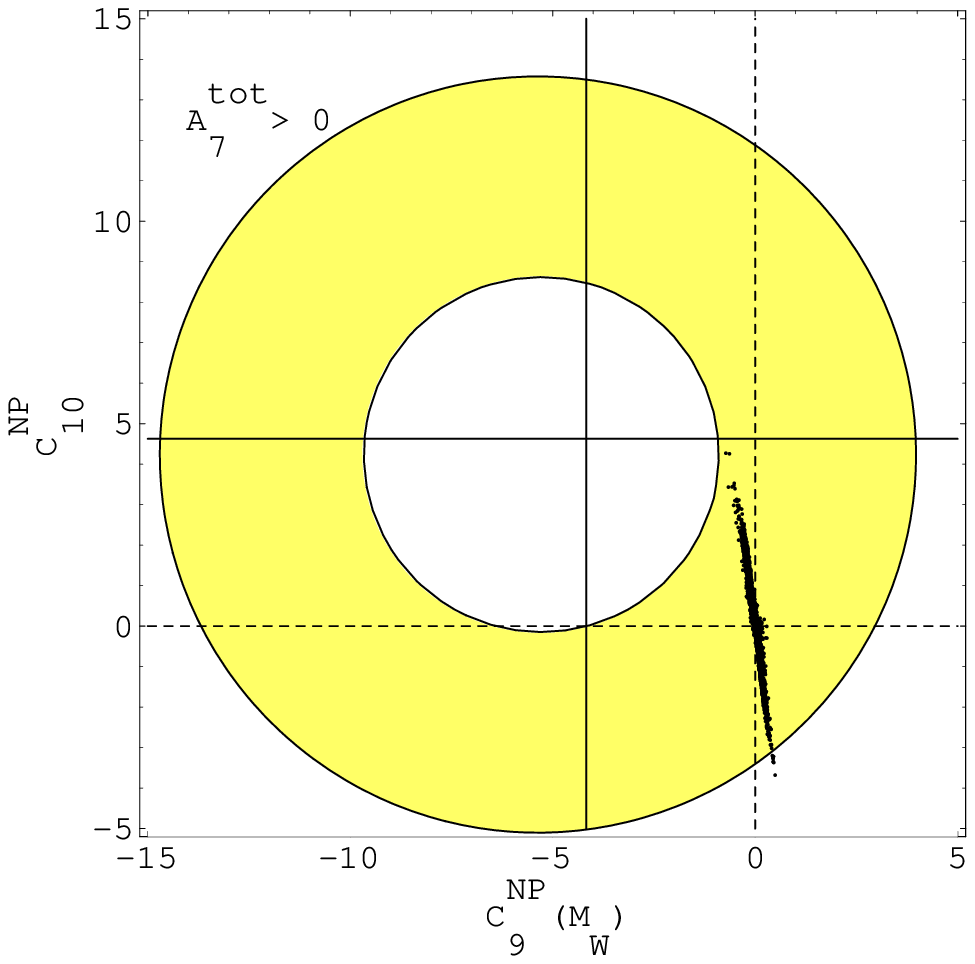,width=0.509\linewidth}
\vskip -0.4truein
\caption{Constraints from $B \to (X_s,K) \ell^+ \ell^-$ data
\cite{Abe:2001dh,Abe:2001qh} on
$C^{NP}_{9}(m_W),C^{NP}_{10}$ for each 
solution allowed by ${\cal{B}}(b \to s \gamma)$ @$90\%$ C.L.,
figure from \cite{aghl}.}
\label{fig:10}
\end{figure}

With data available on $B \to (X_s,K) \ell^+ \ell^-$ decays the model
independent analysis can be extended to include constraints
on NP contributions to 
$C_{9,10}$ for fixed sign of the effective $b s \gamma$ coupling,
see Fig.~\ref{fig:10}. The left plot corresponds to the SM-like sign, 
i.e.~$C_7<0$.
Unlike the $\tan \beta $ enhanced dipole coefficients
$C_{7,8}$ the MSSM-MFV reach in $C_{9,10}$ is small, 
i.e.~the SM is corrected by $\lsim 20 \%$ at $\mu=m_W$.
The scatter plot represents a MSSM scenario with additional
flavor violation, i.e.~mixing between up-squarks of the 
2nd and 3rd generation encoded in
$\delta^U_{23,LL},\delta^U_{23,LR}$.

\begin{figure}[htb]
\vskip -0.2truein
\centerline{\epsfysize=1.6in
{\epsffile{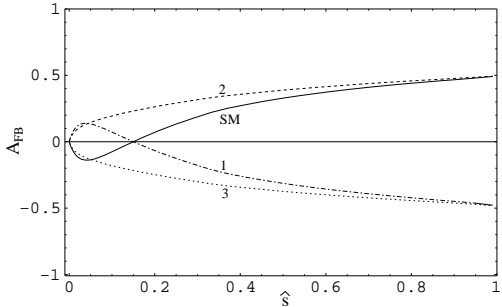}}}
\vskip -0.2truein
\caption{Forward-Backward asymmetry in $b \to s \ell^+ \ell^-$ decays
in the SM (solid) and 3 scenarios beyond the SM as a function of the dilepton
invariant mass. Figure taken from \cite{aghl}.}
\label{fig:afb}
\end{figure}

The Forward-Backward asymmetry  $A_{FB}$  in $b \to s \ell^+ \ell^-$ decays is
defined as
\begin{eqnarray}
\nonumber
A_{FB}(\hat s)\!\! &\!\!\sim\!\!&\!\!
\left (\int_0^1 \!\! d \cos \Theta -\!\!
\int_{-1}^0 \!\! d \cos \Theta \right) 
\frac{d^2 \Gamma}{d \hat s d \cos \Theta} \\
&\sim & -C_{10} \left[ C_7+ \beta (\hat s) Re(C_9)\right]
\nonumber
\end{eqnarray}
where $\Theta$ is the angle between $\ell^+$ and $b$ in the dilepton CMS,
see \cite{GGG} for a discussion of the $A_{FB}$ sign and CP properties.
It tests unique combinations of Wilson coefficients \cite{Ali:1996bm}
and is an ideal NP counter, as illustrated in Fig.~\ref{fig:afb}.
In the SM $A_{FB}$ is negative for very low and
positive for large $\hat s$.
The zero disappears (curve 2) for the non-SM solution $C_7 >0$.
With NP in $C_{10}$ e.g.~induced by non-SM $Z$-penguins \cite{GGG}
the sign of $A_{FB}$ can be flipped (curves 1,3) and also a flat
$A_{FB}(\hat s) \sim 0$ is possible.
The regions labeled in the left plot of Fig.~\ref{fig:10} match the
corresponding numbers and $A_{FB}$ shapes in Fig.~\ref{fig:afb}.
The $A_{FB}$ in exclusive $B \to K^{*0} \mu^+ \mu^-$ decays has analogous
behaviour \cite{ABHH}. The 
expected event yield for $2 fb^{-1}$ at CDF, BTeV, ATLAS, CMS, LHCb is
59, 2240, 665, 4200, 4500 \cite{Eigen:2001tb}, 
respectively and suggests that an experimental
study of $A_{FB}$ is an opportunity for hadron colliders, too.

\section{FLAVOR/CP AND ELECTROWEAK SYMMETRY BREAKING}

\begin{figure}[htb]
\vskip 0.0truein
\centerline{\epsfysize=1.6in
{\epsffile{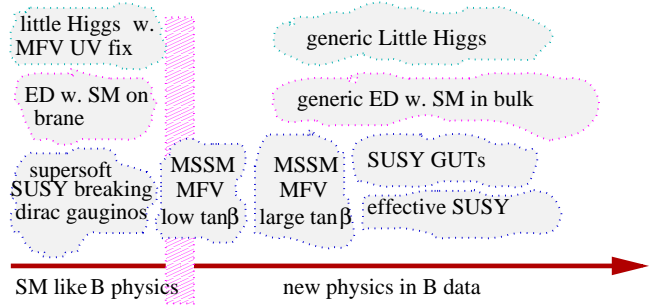}}}
\vskip -0.3truein
\caption{Flavor/CP yield of models of electroweak symmetry breaking.}
\label{fig:survey}
\end{figure}

Realistic extensions of the SM have to address
the hierarchy problem, i.e.~why is the Higgs mass stable against
quadratic corrections arising at 1-loop $\delta m^2_h \sim \Lambda^2/16 \pi^2$
and does not get renormalized up to the Planck scale $\Lambda \sim
10^{19}$GeV ?
Theorists created several frameworks to explain this, which are 
SUSY, models with extra dimensions (ED) 
\cite{Arkani-Hamed:1998rs,Randall:1999ee}, little Higgs (theory
space) \cite{Arkani-Hamed:2002qx} and technicolor  theories plus hybrids.
In all of them we expect to see NP participating in the mechanism of 
electroweak symmetry breaking (EWKSB) 
at/below 1 TeV at the LHC, 
a linear collider or even before at the Tevatron, with technicolor
already being disfavored by precision electroweak data.
Why do we expect to see NP in low energy signals ? This is related to
the question of
how much flavor (and CP) violation is in the model besides the one present
in the SM, i.e.~MFV vs.~non-MFV, as illustrated in the model survey in
Fig.~\ref{fig:survey}.
In the left part are models with SM like $B$-physics, on the right
those with NP in $B$-data. They are separated by a range which size and
position depends on how well we can measure and on the 
theoretical uncertainties.

Flavor violation arises generically in SUSY GUTs
from running above the GUT scale, because of the large yukawas of the 
third generation, e.g.~\cite{Barbieri:1995rs}.
The reported large atmospheric neutrino $\mu-\tau$ 
mixing angle in the context of 
$SO(10)$ embeddings of the MSSM \cite{ESSM,Chang:2002mq}
and extended MSSM \cite{ESSM} has NP consequences for rare processes,
in particular for the $b$-system.
Effective SUSY with first 2 generations of sfermions heavy 
but the third below a TeV \cite{Cohen:1996sq,Hisano:2000wy} does 
predict NP effects in $B$-data. 
The supersoft proposal \cite{Fox:2002bu} with 
all squarks above 1 TeV and highly
degenerate SUSY breaking however will escape low energy searches.
In ED scenarios the flavor yield 
is model dependent, i.e.~depends on the location of
the SM fields: if they -or part of the SM- live in the bulk, new 
sources of flavor changing neutral currents (FCNC) arise
\cite{Delgado:1999sv}.
Generic little Higgs models do have anomalous top couplings which can
yield interesting flavor physics because of the low
cut off $\Lambda \sim 10$ TeV, see \cite{Chivukula:2002ww}.
This might be evaded by a
clever choice of UV completion \cite{Arkani-Hamed:2002qx}.
All above EWKSB frameworks can lead to NP signals in rare decays. 
Hence, experimental study at the $b$-factories and the Tevatron
-if we are lucky even before the LHC- can establish NP and 
as shown in Sec.~\ref{sec:pattern} distinguish between models.
Particularly interesting (experimentally feasible, theoretically clean
SM interpretation) observables are the top10
beyond ${\cal{B}}(b \to s \gamma)$ given in the next Sec.~\ref{sec:top10},
which probe different aspects of the underlying theory, 
e.g.~CP, $b s g$, chirality.

\section{TOP10 OBSERVABLES SEEKING NP \label{sec:top10}}

1.~The CP asymmetry in $b \to s \gamma$ decays. In the 
SM direct CP violation in $b \to s$ transitions is small
$a_{CP}=\frac{|A|^2-|\bar A|^2}{|A|^2+|\bar A|^2} \sim \alpha_s(m_b) 
Im \frac{V_{ub} V_{us}^*}{V_{tb} V_{ts}^*} \sim \alpha_s(m_b) \lambda^2\lsim
{\cal{O}}(1 \%) $, e.g.~\cite{hillerkagan}. 
This is experimentally probed at the 10 $\%$ level
$a_{CP}=(-0.079 \pm 0.108 \pm 0.022)(1 \pm0.03)$ by
Cleo \cite{Coan:2000pu}.
2.~Wrong helicity contributions to
$\bar{s}_R \sigma_{\mu \nu} b_L F^{\mu \nu}$ in $b \to s \gamma$
decays. In the SM
this is small $C_7^\prime=m_s/m_b C_7$. It can be tested 
e.g.~with polarization studies in
$\Lambda_b \to (\Lambda \to p \pi)  \gamma$ at hadron colliders and 
GigaZ \cite{hillerkagan}. 
3.~Time dependent study in $B, \bar B \to \Phi K_{S,L}$ decays.
The difference 
$|\sin 2 \beta_{(J/\Psi K)}-\sin 2 \beta_{(\Phi K)}|$ is $\lsim
{\cal{O}}(\lambda^2)$ in the SM and probes 
direct CP violation in $b \to s \bar s s$ decays even in the absence
of strong phases.
The precision expected at the $B$-factories is
$\sigma_{\Phi K_S}(stat)=0.56,0.18$ for $0.1,1ab^{-1}$ \cite{Eigen:2001mk}.
4.~Precision study of the inclusive $b \to s \ell^+ \ell^-$ 
branching ratio for low $q^2$ below the charm threshold \cite{aghl}.
5.~Sign and shape of $A_{FB}(B \to (X_s,K^*) \ell^+ \ell^-)$ \cite{aghl,ABHH}.
6.~If it exists, the position of the $A_{FB}$ zero \cite{ABHH}.
7.~The Forward-Backward-CP asymmetry 
$A_{FB}^{CP} \equiv \frac{A_{FB}+\bar{A}_{FB}}{A_{FB}-\bar{A}_{FB}}\sim
\frac{Im(C_{10})}{Re(C_{10})}$ above the $\Psi^\prime$ to have
sizeable strong phase
probes non-SM CP violation in $C_{10}$. The SM background is tiny
$A_{FB}^{CP}< 10^{-3}$ \cite{GGG}.
8.~$B_s-\bar B_s$ mixing.
9.~${\cal{B}}(B_{d,s} \to \mu^+ \mu^-)$ is sensitive to neutral Higgs exchange
\cite{Bobeth:2001sq}.
10.~The non-observation of nucleon electric dipole moments (nEDMs)
created the strong CP problem, i.e.~why is 
$\bar \Theta < 10^{-10}$ while $\delta_{CKM} \sim O(1)$ ?
nEDMs are sensitive to flavor blind CP 
violation in case of the PQ-axion solution. In models with
spontaneously broken CP tight constraints on the flavor structure
arise, suggesting that nEDMs could be close to the
current bounds \cite{Hiller:2002um}.

\section{NEW PHYSICS PATTERN \label{sec:pattern} }

If the SM is extended by adding either more a) symmetry,  
b)  Higgs, c) matter or d) gauge interactions
distinct pattern of NP signals in low energy observables arise. 
This is illustrated in Tab.~\ref{tab:pattern} 
for the MSSM with MFV \cite{ABHH,Bobeth:2001sq}, 
the 2 Higgs doublet model (2HDM) III \cite{Bowser-Chao:1998yp}, 
which contains an extra source of
CP violation, a model with a vector-like down quark (VLdQ)
\cite{Ahmady:2001qh} and
anomalous top couplings \cite{Burdman:1999fw}.
All except the MSSM are toy models, but they can be part of a
complete model of EWKSB and mimik 
e.g.~the enlarged Higgs sector of little Higgs theories or the
Kaluza-Klein states in models with EDs.
Note that a non-SM $sZb$ vertex includes NP in 
$A_{FB},A_{FB}^{CP}$, $b \to s \nu \bar \nu$ decays
and $B_s-\bar B_s$ mixing \cite{GGG}.
\begin{table*}[htb]
\caption{New physics pattern in $b$-decay observables.
The ! indicates drastic non-SM effect possible.}
\label{tab:pattern}
\renewcommand{\arraystretch}{1.12}
         \begin{tabular}{l|c|c|c|c|c|c}
\hline
example   
& flip $C_7$ sign& $a_{CP}^{b \to s \gamma}$ & 
$\arg(\frac{\bar A}{A})_{\Phi K_{S,L}}$ &
         $C_7^\prime$& $sZb $  & $B \to \mu^+ \mu^-$\\
\hline
a) MSSM+MFV & ! &$\lsim 1 \%$ &--&$\sim \frac{m_s}{m_b}$&$\lsim 20 \%$
& ! for large $\tan \beta$
\\
b) 2HDM III &-- & $\lsim 1 \%$& $ ! $ &$\sim \frac{m_s}{m_b}$ &--&--
\\
c) VLdQ&-- &$\lsim 1 \%$ & $ !  $ &$\sim
         \frac{m_s}{m_b}$ & !& --
\\
d) anomal.~coupl. &!&!&!&!&!&!\\
\hline
         \end{tabular}
\end{table*}
Also in the last row of Tab.~\ref{tab:pattern} lives the 
MSSM without R parity \cite{Harrison:1998yr}
and the MSSM with generic soft terms.
A direct (non-FCNC) determination of
$V_{tq}$ is important since 
extra quarks generally violate CKM unitarity
$\sum_{i=u,c,t} \! V_{i b} V_{iq}^* \! \neq 0$, $q=d,s$.

\section{SUMMARY}

Running and upcoming $b$-facilities allow for an extension of the program of 
FCNC tests which started a decade ago with 
$B \to K^* \gamma $ decays. 
Further flavor/CP sensitive observables 
are in reach now
and rare $b \to s \ell^+ \ell^-$ decays have begun to be measured. They
complement the radiative modes and
collider searches, with
particularly clean observables 
${\cal{B}}(B \to X_s \ell^+ \ell^-)$
for low dilepton mass and the Forward-Backward-asymmetry in inclusive and
exclusive $B \to K^* \ell^+ \ell^- $ decays (Fig.~\ref{fig:afb}).

The analysis presented here to search for NP in the
short distance coefficients $C_{7,8,9,10}$ is only model independent
as long as the operator basis in the effective Hamiltonian is complete. 
Certain observables listed in the top10 such as
wrong helicity contributions require additional operators beyond those
present in the SM.
With NP at a TeV as suggested by models of EWKSB there is a good chance  
(Fig.~\ref{fig:survey}) that it will show up in one or the other observables
@ 5GeV.
In analogy with the determination of
the parameters of the CKM matrix
only a global analysis of all FCNC and low energy data might reveal
NP and whether it violates also CP. This procedure is 
able to distinguish between models (Tab.~\ref{tab:pattern}).
Constraints from 
rare $K,D$ and lepton flavor violating processes and
neutrino physics complete the picture.

\noindent
{\bf ACKNOWLEDGEMENTS}
It is a pleasure to thank the organizers for the invitation to 
Vancouver. I would like to thank 
Ahmed Ali, Christoph Greub and Enrico Lunghi for collaboration \cite{aghl}
and 
Gustavo Burdman, David E.~Kaplan, Jogesh Pati, Tom Rizzo,
Martin Schmaltz and Stefan Spanier for communication.


\begin{thebibliography}{9}
\bibitem{misiak}
P.~Gambino and M.~Misiak,
Nucl.\ Phys.\ B {\bf 611}, 338 (2001), hep-ph/0104034, and references therein.


\bibitem{aghl} A.~Ali {\it et al.},
hep-ph/0112300, and references therein.

\bibitem{Eigen:2001tb}
G.Eigen,
{\it Snowmass 2001},
hep-ex/0112041.

\bibitem{Abe:2001dh}
K.~Abe {\it et al.}  [BELLE],
Phys.\ Rev.\ Lett.\  {\bf 88}, 021801 (2002),
hep-ex/0109026.

\bibitem{babarKll}
J.~Walsh, talk presented at FPCP, 16-18 May 2002, Philadelphia.

\bibitem{ABHH}
A.~Ali {\it et al.}, 
Phys.\ Rev.\ D {\bf 61}, 074024 (2000),
hep-ph/9910221, and references therein.

\bibitem{BobethNNLO}
C.~Bobeth, M.~Misiak and J.~Urban,
Nucl.\ Phys.\ B {\bf 574}, 291 (2000),
hep-ph/9910220.


\bibitem{Asatryan:2001zw}
H.~H.~Asatryan {\it et al.},
Phys.\ Rev.\ D {\bf 65}, 074004 (2002), 
hep-ph/0109140.


\bibitem{Asatryan:2002iy}
H.~H.~Asatryan {\it et al.},
hep-ph/0204341.

\bibitem{Buchalla:1997ky}
G.~Buchalla, G.~Isidori and S.~J.~Rey,
Nucl.\ Phys.\ B {\bf 511}, 594 (1998),
hep-ph/9705253.


\bibitem{Ali:1996bm}
A.~Ali {\it et al.}, 
Phys.\ Rev.\ D {\bf 55}, 4105 (1997),
hep-ph/9609449.

\bibitem{Abe:2001qh}
K.~Abe {\it et al.}  [BELLE],
hep-ex/0107072.

\bibitem{bellesenyo}
K.~Senyo, talk presented at FPCP, 
16-18 May 2002, Philadelphia, hep-ex/0207005.

\bibitem{GGG}
G.~Buchalla, G.~Hiller and G.~Isidori,
Phys.\ Rev.\ D {\bf 63}, 014015 (2001),
hep-ph/0006136, and references therein.


\bibitem{Arkani-Hamed:1998rs}
N.~Arkani-Hamed, S.~Dimopoulos and G.~R.~Dvali,
Phys.\ Lett.\ B {\bf 429}, 263 (1998),
hep-ph/9803315.

\bibitem{Randall:1999ee}
L.~Randall and R.~Sundrum,
Phys.\ Rev.\ Lett.\  {\bf 83}, 3370 (1999),
hep-ph/9905221.

\bibitem{Arkani-Hamed:2002qx}
N.~Arkani-Hamed {\it et al.},
hep-ph/0206020, and references therein.



\bibitem{Barbieri:1995rs}
R.~Barbieri, L.~J.~Hall and A.~Strumia,
Nucl.\ Phys.\ B {\bf 449}, 437 (1995),
hep-ph/9504373.

\bibitem{ESSM} K.S.Babu and
J.C.Pati,
To appear and Private communications; hep-ph/0203029.

\bibitem{Chang:2002mq}
D.~Chang, A.~Masiero and H.~Murayama,
hep-ph/0205111.

\bibitem{Cohen:1996sq}
A.~G.~Cohen {\it et al.},
Phys.\ Rev.\ Lett.\  {\bf 78}, 2300 (1997), 
hep-ph/9610252.

\bibitem{Hisano:2000wy}
J.~Hisano, K.~Kurosawa and Y.~Nomura,
Nucl.\ Phys.\ B {\bf 584}, 3 (2000),
hep-ph/0002286.

\bibitem{Fox:2002bu}
P.~J.~Fox, A.~E.~Nelson and N.~Weiner,
hep-ph/0206096.

\bibitem{Delgado:1999sv}
A.~Delgado, A.~Pomarol and M.~Quiros,
JHEP {\bf 0001}, 030 (2000),
hep-ph/9911252.

\bibitem{Chivukula:2002ww}
R.S.~Chivukula, N.~Evans and E.H.~Simmons,
hep-ph/0204193.

\bibitem{hillerkagan}
G.~Hiller and A.~Kagan,
Phys.\ Rev.\ D {\bf 65}, 074038 (2002),
hep-ph/0108074, and references therein.

\bibitem{Coan:2000pu}
T.~E.~Coan {\it et al.}  [CLEO],
Phys.\ Rev.\ Lett.\  {\bf 86}, 5661 (2001),
hep-ex/0010075.

\bibitem{Eigen:2001mk}
G.Eigen {\it et al.}, 
{\it Snowmass01},
hep-ph/0112312.


\bibitem{Bobeth:2001sq}
C.~Bobeth {\it et al.}, 
Phys.\ Rev.\ D {\bf 64}, 074014 (2001),
hep-ph/0104284.

\bibitem{Hiller:2002um}
G.~Hiller and M.~Schmaltz,
Phys.\ Rev.\ D {\bf 65}, 096009 (2002),
hep-ph/0201251, and references therein.


\bibitem{Bowser-Chao:1998yp}
D.~Bowser-Chao, K.~m.~Cheung and W.~Y.~Keung,
Phys.\ Rev.\ D {\bf 59}, 115006 (1999),
hep-ph/9811235.

\bibitem{Ahmady:2001qh}
M.~R.~Ahmady, M.~Nagashima and A.~Sugamoto,
Phys.\ Rev.\ D {\bf 64}, 054011 (2001),
hep-ph/0105049.

\bibitem{Burdman:1999fw}
G.~Burdman, M.~C.~Gonzalez-Garcia and S.~F.~Novaes,
Phys.\ Rev.\ D {\bf 61}, 114016 (2000),
hep-ph/9906329.

\bibitem{Harrison:1998yr}
``can do everything except make coffee'', 
P.~Harrison and H.~Quinn,
SLAC-R-0504.
\end{thebibliography}
\end{document}